\documentclass[aps,reprint,superscriptaddress]{revtex4-1}

\usepackage{amsmath}    
\usepackage{graphicx}   
\usepackage{xcolor}      

\usepackage{nicefrac}
\usepackage[latin1]{inputenc}

\usepackage{hyperref}   

\begin{document}

\title{Challenging the nature of low energy plasmon excitations in \texorpdfstring{CaC$_6$}{CaC6} using electron energy-loss spectroscopy}


\author{Friedrich Roth}
\author{Andreas K\"onig}
\affiliation{IFW Dresden, P.O. Box 270116, D-01171 Dresden, Germany}

\author{Christian Kramberger}
\author{Thomas Pichler}
\affiliation{University of Vienna, Faculty of Physics, Strudlhofgasse 4, AT-1090, Vienna, Austria}

\author{Bernd B\"uchner}
\author{Martin Knupfer}
\affiliation{IFW Dresden, P.O. Box 270116, D-01171 Dresden, Germany}


\date{\today}

\begin{abstract}
The nature of low energy plasmon excitations plays an important role in understanding the low energy electronic properties and coupling mechanism of different superconducting compounds such as CaC$_6$. Recent \emph{ab-initio} studies predict a charge carrier intraband plasmon in keeping with a low energy acoustic plasmon. Here, we have studied the low-energy electronic excitations of CaC$_6$ using high-resolution electron energy-loss spectroscopy in transmission at low temperatures. The analysis of the core-level excitations leads to the conclusion that hybridization between graphite and calcium states plays an essential role in this graphite intercalated compound. Regarding the low energy plasmon excitation, we observe the formation of an intraband (charge carrier) plasmon with a negative dispersion at about 3.5\,eV in sound agreement with the theory.
Finally, a weak excitation around 1.2\,eV with an almost linear dispersion relation can be observed as predicted for an acoustic plasmon that may mediate the superconducting coupling in CaC$_6$. However its optical limit at $\sim$~1\,eV challenges the theoretical predictions and safely rules out an electronic superconducting coupling mechanism in CaC$_6$.
\end{abstract}

 \maketitle

\section{Introduction}

The physical properties of graphite intercalation compounds have been in the focus of a number of research activities since many years \cite{Dresselhaus1981}. Such compounds are synthesized by the insertion of particular atoms or molecules in-between the individual layers of graphite in order to tune the physical properties. In many cases, the dopants provide graphite with charge carriers, and the original semimetal can be transformed into a good metal with a substantially increased density of states at the Fermi level \cite{Dresselhaus1981,Oelhafen1980,Mizutani1978}. In addition, doping of graphite can be achieved with holes or electrons depending on the intercalant that is provided \cite{Ritsko1979,Mele1979,Grunes1983}.

\par

One of the fascinating properties that can be induced by intercalation of graphite is the appearance of superconductivity. Prominent examples are the alkali metal intercalation compounds KC$_8$, RbC$_8$, and CsC$_8$ with transition temperatures into the superconducting state below 0.5\,K \cite{Koike1980,Hannay1965}. Subsequently, several ternary graphite intercalation compounds revealed higher $T_c$'s of 1.4\,K for KHgC$_8$ \cite{Pendrys1981} and 2.7\,K for KTl$_{1.5}$C$_4$ \cite{Wachnik1982} and further the transition temperature was increased up to 5\,K by changing the dopant as well as other parameters such as pressure \cite{Belash1987,Belash1989,Belash1990,Belash1990_2}.

\begin{figure}[t]
\centering
\includegraphics[width=0.7\linewidth]{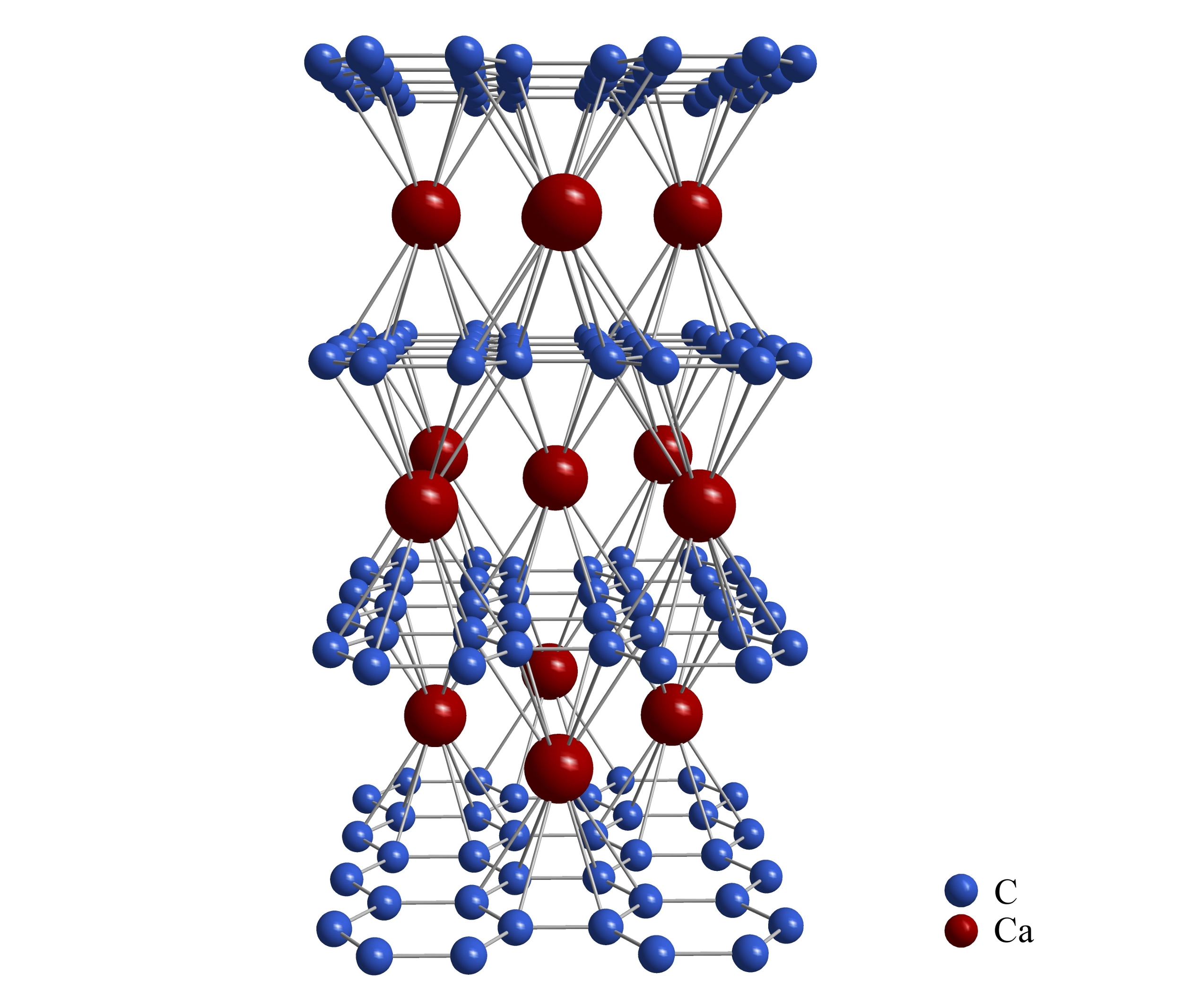}
\caption{Crystal structure of CaC$_6$, whereby the positions of the carbon atoms are shown by small blue balls and those of calcium ions by large red balls.} \label{f1}
\end{figure}

Recently, the discovery of a superconducting ground state in CaC$_6$ (cf.\,Fig.\,\ref{f1}) at a relatively high temperature ($T_c$ $\sim$ 11.5\,K) refocused research activities on graphite intercalation compounds \cite{Weller2005,Emery2005}. In particular the mechanism that is responsible for the pairing of electrons in the superconducting state has been discussed. While it is now generally accepted that electron phonon coupling can explain the relatively high transition temperature in CaC$_6$ \cite{Calandra2006,Takada2009,Calandra2005,Lamura2006}, there are some open issues in this context \cite{Mazin2007}. Based on the fact that in CaC$_6$ different bands are filled with electrons and cross the Fermi level, it has even been suggested that so-called acoustic plasmons might exist and perhaps are able to mediate pairing  \cite{Geilikman1966,Ruvalds1981,Wohlman1984,Mazin2007}.

\par

Plasmons, the collective electronic excitations of the charge carriers in metals, have also been studied in graphite intercalation compounds. It has been demonstrated that the intercalation with alkali metals or FeCl$_3$ and the respective charge transfer results in charge carrier plasmons at about 2.3\,eV (for e.\,g. KC$_8$) and 1.1\,eV (for FeCl$_3$) with a quadratic dispersion as would be expected for (Drude-like) metallic systems \cite{Ritsko1982,Hwang1981,Hwang1979,Mele1980}.

For CaC$_6$, the energy-loss function $\operatorname{Im}$[-1/$\epsilon(q,\omega)$] has recently been calculated as a function of
different momentum values and directions \cite{Echeverry2012}. Interesting results have been reported, among them an intraband plasmon with a negative dispersion and the appearance of a low energy acoustic plasmon. The existence of acoustic plasmons in two-component electron systems with vastly different masses or with a significant two-dimensional character has been suggested decades ago \cite{Froehlich1968,Wohlman1984,Schaefer1987,Kresin1988}, and it has been argued that such plasmons might exist in Pd metal and MgB$_2$ \cite{Silkin2009,Silkin2009_2}. For CaC$_6$, the compound discussed here, however also arguments against the presence of acoustic plasmons have been put forward \cite{Mazin2005}. 

\par

In this contribution we report experimental studies of the electron-loss function of CaC$_6$ as a function of momentum within the hexagonal $a$,$b$ crystal plane. The loss function has been measured using electron energy-loss spectroscopy (EELS) in transmission. EELS measurements in the past already have been carried out on materials based on $sp^2$-hybridized carbon \cite{Knupfer2001,Kramberger2011}, its intercalation compounds \cite{Liu2003,Liu2004_2} and related systems, and they have provided important insight into the nature of the electronic excitations, their degree of delocalization, the importance of crystal local field effects and the particular band structure \cite{Marinopoulos2002,Mele1979,Ritsko1981}. We demonstrate that the electronic excitation spectrum significantly changes upon Ca addition to graphite, and that there are two new collective excitations below the well known $\pi$ plasmon. The character and dispersion of these two excitations is discussed.

\section{Experimental}

The preparation of Ca intercalated graphite has been carried out using highly orientated pyrolytic graphite (HOPG) platelets purchased from
MaTecK (MaTecK GmbH, Germany). For the EELS measurements, thin films ($\sim$\,100\,nm thickness) are required. These were cut perpendicular to the crystallographic $c$-axis from the HOPG platelets using an ultramicrotome equipped with a diamond knife. Subsequently, the films were put onto standard transmission electron microscopy grids and transferred into the spectrometer. Prior to the EELS measurements the films were characterized \emph{in situ} using electron diffraction, in order to orient the crystallographic axis with respect to the transferred momentum. All observed diffraction peaks were consistent with the crystal structure of pristine graphite \cite{TRUCANO1975}.

\par

Ca intercalation was achieved by evaporation of Ca from a commercial molybdenum crucible (Omicron Nano\-Technology GmbH, Germany) onto the thin graphite films under ultra-high vacuum conditions (base pressure lower than 10$^{-10}$\,mbar). This has been carried out in several steps until saturation, i.\,e., until no further stoichiometry change could be observed from the spectra. During Ca evaporation, the films were kept at room temperature, then they were post-annealed at 300\,$^\circ$C for several hours in every preparation step. This was done to achieve a homogeneous Ca distribution in the films and removed calcium overlayers on the film surfaces. The film composition has been analyzed via the analysis of C\,1$s$ and Ca\,2$p$ core level intensities as described in more detail below. This analysis demonstrated that the saturation composition was CaC$_6$. We estimate the error of this analysis to be about 15\,\%.

\par

All loss function and core excitation measurements were carried out with a dedicated transmission electron energy-loss spectrometer with a
primary electron energy of 172\,keV \cite{Fink1989}. We note that at this high primary beam energy only singlet excitations are possible. The energy and momentum resolution were chosen to be 85\,meV and 0.03\,\AA$^{-1}$, for the loss function measurements and 360\,meV and
0.03\,\AA$^{-1}$ for the core excitations, respectively. We have measured the loss function $\operatorname{Im}$[-1/$\epsilon(q,\omega)$], which is proportional to the dynamic structure factor S($q,\omega$), for a momentum transfer $q$ parallel to the film surface [$\epsilon(q,\omega)$ is the dielectric function], i.\,e., for a momentum transfer within the $a,b$ crystallographic plane. During the measurements the samples were cooled down to 20\,K.

\section{Results and discussion}

\begin{figure}[ht]
\centering
\includegraphics[width=.87\linewidth]{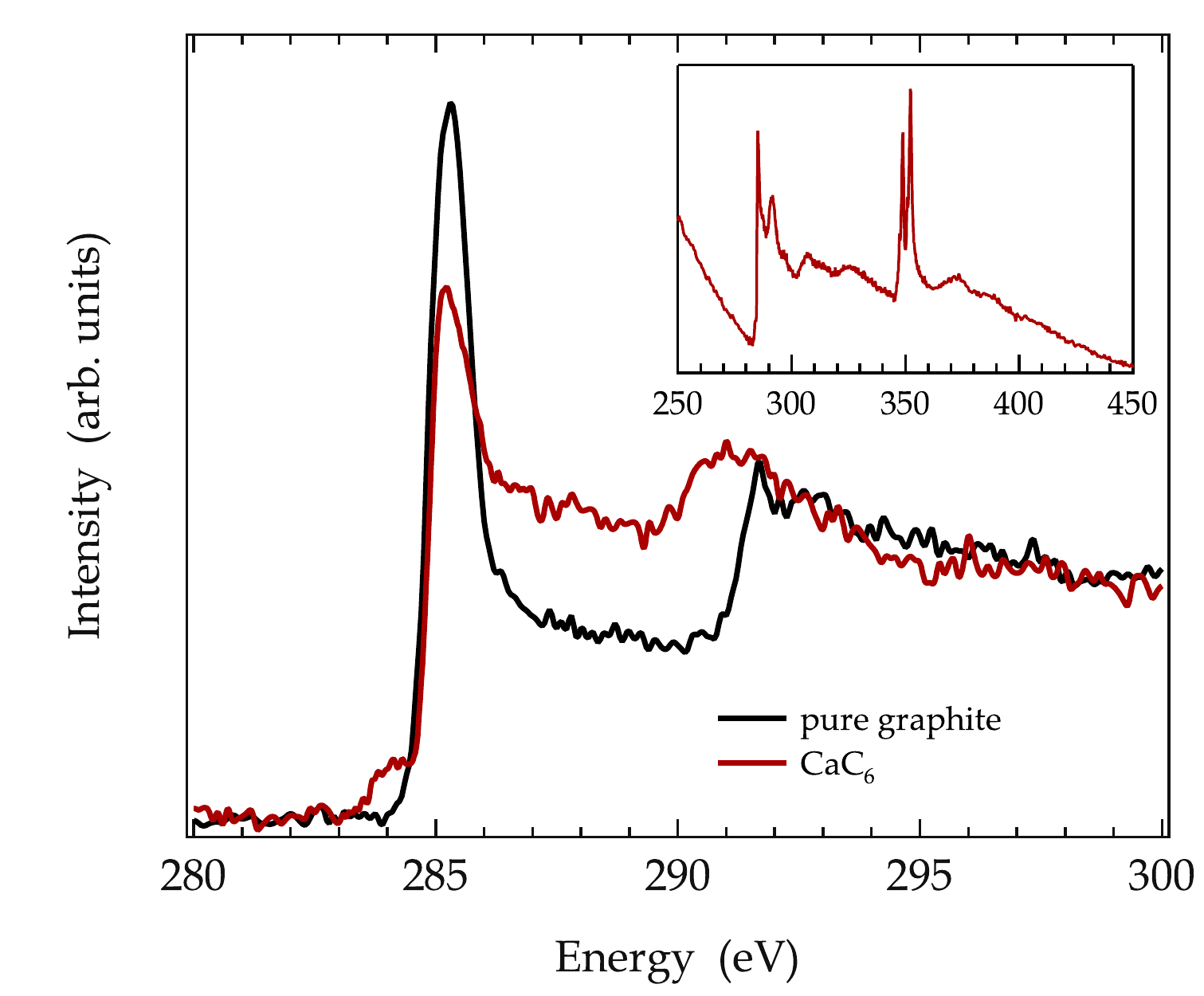}
\caption{C\,1$s$ core-level excitations of pristine and calcium doped graphite. Inset: Core level excitation profile of calcium intercalated graphite for a wide energy range between 200 and 450\,eV, where the C\,1$s$ and the Ca\,2$p$ core excitations are visible.} \label{f2}
\end{figure}

We start the discussion of the doping introduced changes in the electronic structure of graphite upon Ca addition with the analysis of the C\,1$s$ and Ca\,2$p$ core level excitations. In Fig.\,\ref{f2} we show the C\,1$s$ as well as Ca\,2$p$ core-level excitations for calcium doped graphite, and also the C\,1$s$ data of pure graphite for comparison. First, these data can be used to determine the stoichiometry of the film under investigation by analyzing the relative core level excitation intensities of the C\,1$s$ and Ca\,2$p$ core excitation edges. We have subtracted the background following the procedures described in \cite{Egerton1989,Hofer1991,Egertonbuch} and integrated the intensities of the C\,1$s$ and Ca\,2$p$ excitation over an energy range of 62 and 100\,eV, respectively. The atomic ratio has been obtained by weighting the experimental intensities with calculated cross sections using the SIGMAK and SIGAML progams \cite{Egertonbuch}. We estimate the error of this procedure to be about 15\,\% and obtain a composition for our films of CaC$_6$.

Moreover, equivalent to other $\pi$ conjugated materials \cite{Knupfer2001,Roth2010,Roth2010_2}, the C\,1$s$ excitation (cf.~Fig.\,\ref{f2}) represents excitations into the carbon 2$p$ contribution to unoccupied $\pi/\pi^*$-states, and thus allows to probe the projected unoccupied electronic density of states of carbon-based materials. In Fig.\,\ref{f2} we depict a comparison of the carbon C\,1$s$ excitation of pure, as well as calcium intercalated graphite. The spectra were normalized at the high energy tail. For the undoped graphite we can clearly identify a sharp and strong feature around 285.3\,eV, which can be assigned to transitions into $\pi^*$-states. These data are in very good agreement to previous publications \cite{Batson1993,Knupfer1999,Kuzuo1992}. Interestingly, upon addition of calcium, a reduction of the spectral weight of the C\,1$s$ excitation can be observed, which signals a successful doping of the film.

Furthermore, the data allow the identification of a new spectral feature at lower excitation energy (around 284\,eV) in the case of CaC$_6$. We note that such a lower energy feature is not observed for potassium intercalated graphite materials. Also, an increase of spectral weight in the region between 286 - 290\,eV in case of calcium intercalation is observable. These changes can be explained by a hybridization of carbon and alkali-metal derived states, which are present in CaC$_6$. Interestingly, similar results, as obtained in our case for Ca intercalated graphite, were observed for Ba intercalated single-wall carbon nanotubes (SWCNT) \cite{Liu2004}. Analogously, for the Ba intercalated nanotubes it was argued that there are hybridized states of Ba and SWCNT. Consequently, our core level excitation data demonstrate the achieved sample composition, and the formation of hybrid bands as predicted by density functional based calculations \cite{Csanyi2005,Echeverry2012}.

\par

\begin{figure}[h]
\centering
\includegraphics[width=.87\linewidth]{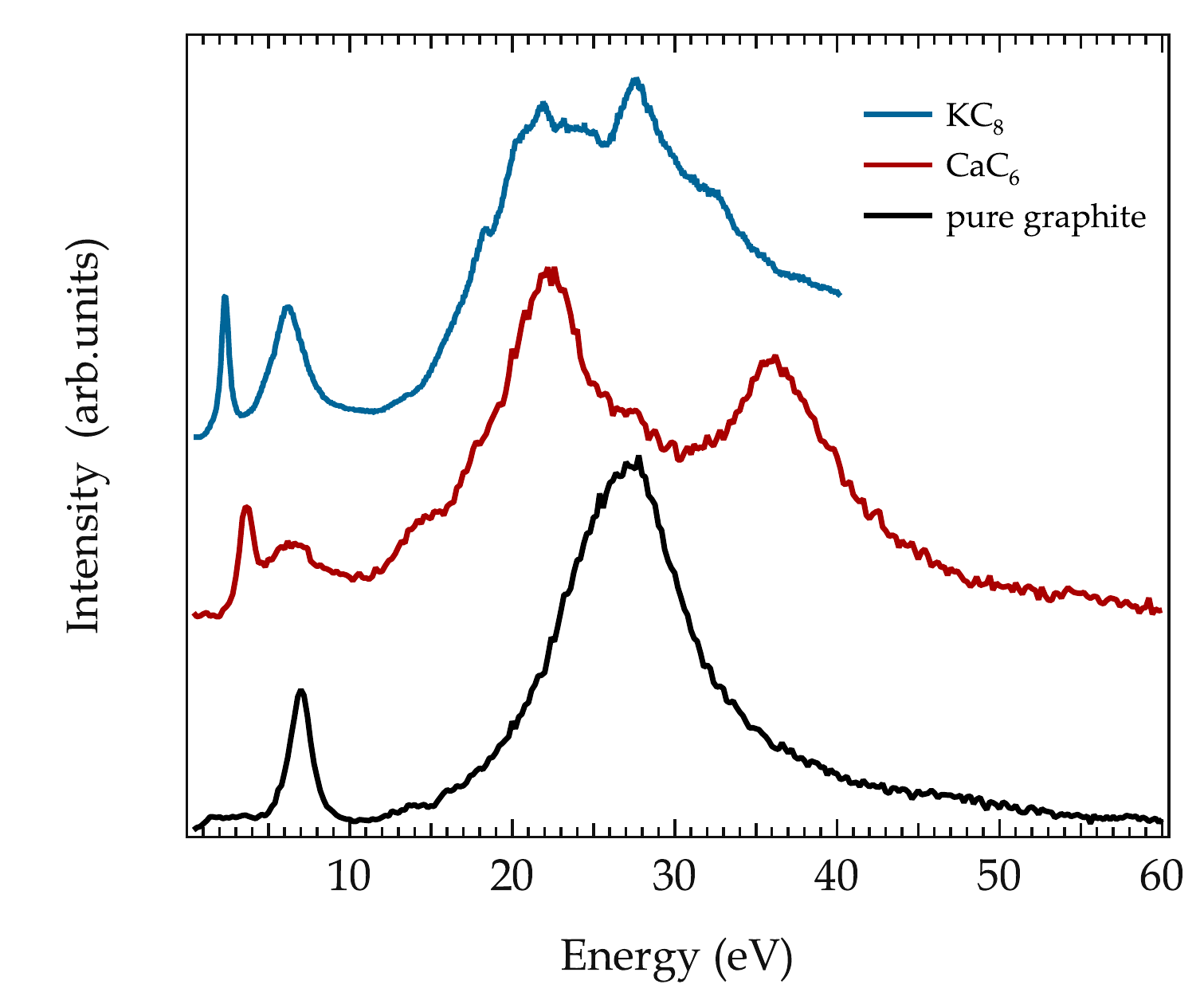}
\caption{Comparison between the loss function of pure graphite (lower spectrum), calcium doped graphite (middle spectrum), and potassium doped graphite (upper spectrum) in an energy range between 0 and 60\,eV. (For additional informations and details of the doping procedure in case of KC$_8$ see Ref.\,\citenum{Pichler1999})
} \label{f3}
\end{figure}

In Fig.\,\ref{f3} we present the loss function for the undoped, as well as calcium (CaC$_6$), and potassium (KC$_8$) intercalated graphite measured at 0.1\,\AA$^{-1}$ momentum transfer in an wide energy range between 0 and 60\,eV. At such a small momentum transfer, we probe the so-called optical limit or excitations close to the center of the Brillouin zone. The spectrum of pristine graphite is dominated by a broad peak at around 27\,eV, which corresponds to the $\pi+\sigma$ plasmon and is a result of the collective oscillation of all valence electrons. Furthermore, one can identify a well defined peak at 7\,eV corresponding to the $\pi$ plasmon related to the oscillation of all $\pi$ electrons in the system. These observations are in common with previous optical \cite{Klucker1974,Taft1965} and EELS measurements \cite{Zeppenfeld1968,Zeppenfeld1971,Buechner1977}. The situation clearly changes when graphite is doped via the addition of alkali or alkaline earth metals. As shown in Fig.\,\ref{f3} for both graphite intercalation compounds, the volume plasmon shifts to lower energies. In addition, new spectral features can be observed. In case of KC$_8$ the complex structure above 15\,eV has previously been explained as a sum of potassium 3$p$ core excitations, graphite interband transitions in the folded Brillouin zone imposed by the potassium superlattice, and the volume plasmon \cite{Ritsko1981,Grunes1983}. Below 10\,eV one can identify a new maximum below the $\pi$ plasmon, at about 2.3\,eV, which corresponds to the charge carrier plasmon in this compound \cite{Grunes1983}.

\begin{figure}[t]
\centering
\includegraphics[width=.87\linewidth]{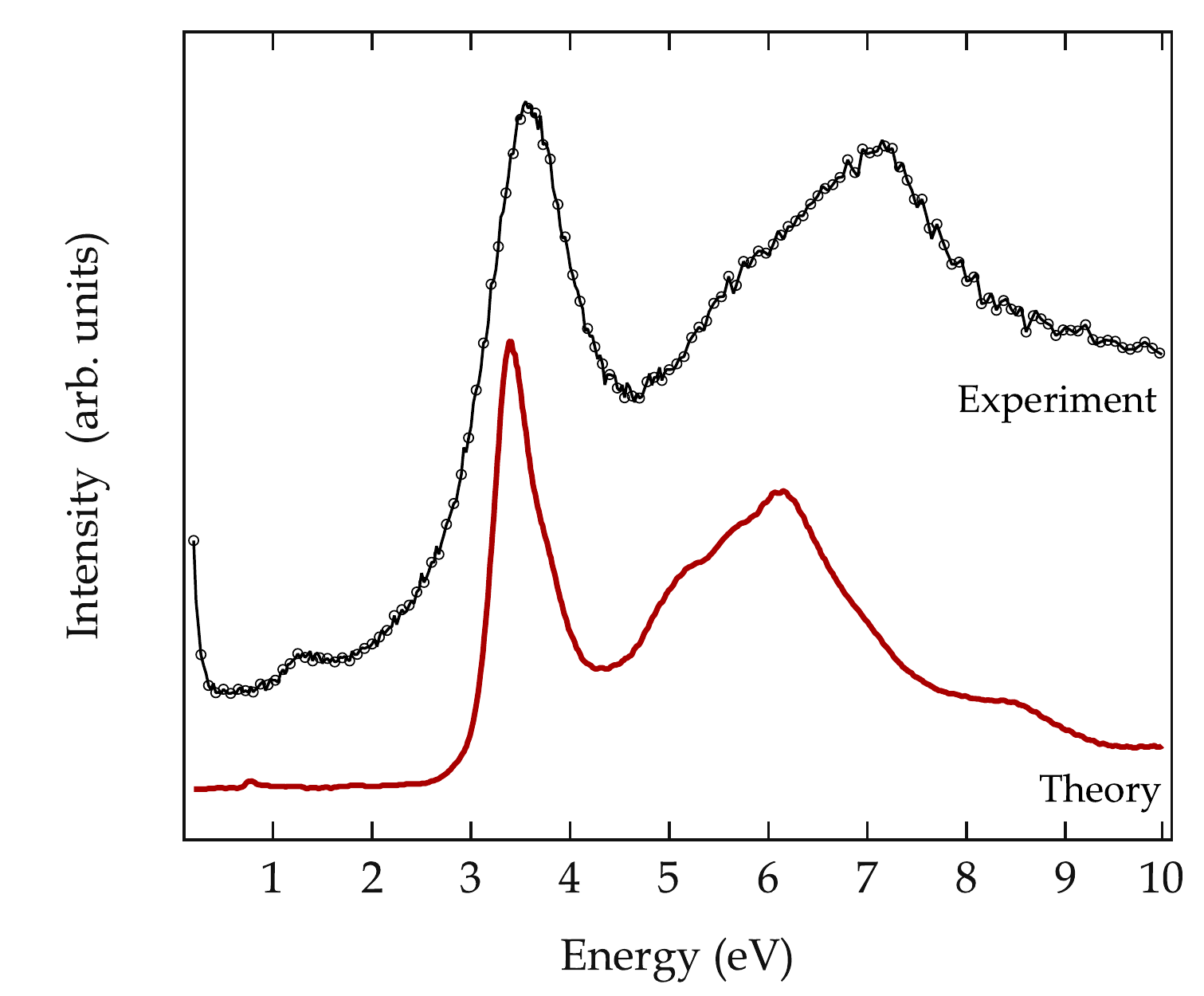}
\caption{Comparison between the measured loss function with a momentum transfer of $q$\,=\,0.175\,\AA$^{-1}$ and the calculated loss function for $q$\,=\,0.181\,\AA$^{-1}$ taken from \cite{Echeverry2012}.
} \label{f4}
\end{figure}

\begin{figure*}[ht]
\centering
\includegraphics[height=6.5cm]{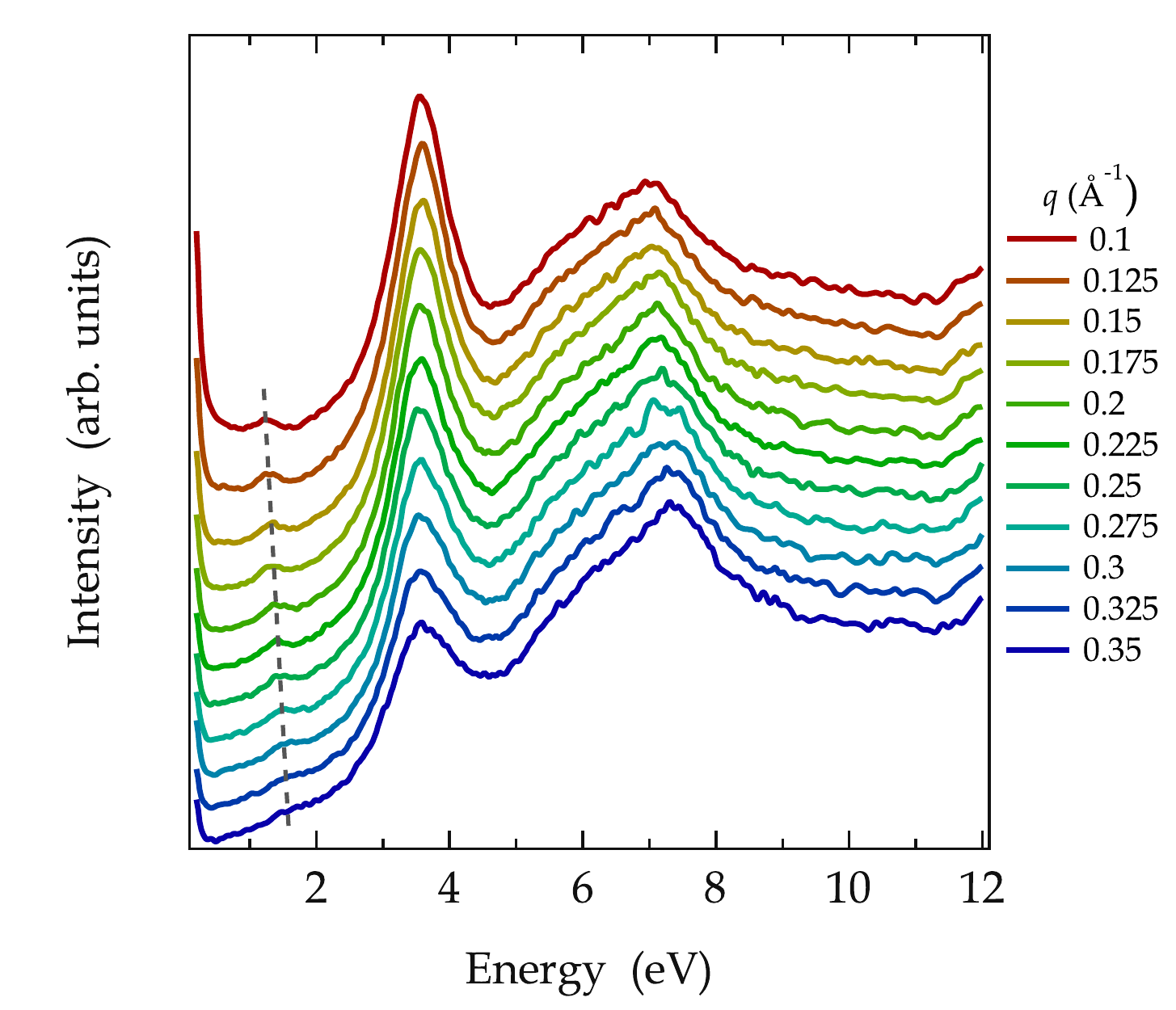}
\includegraphics[height=6.5cm]{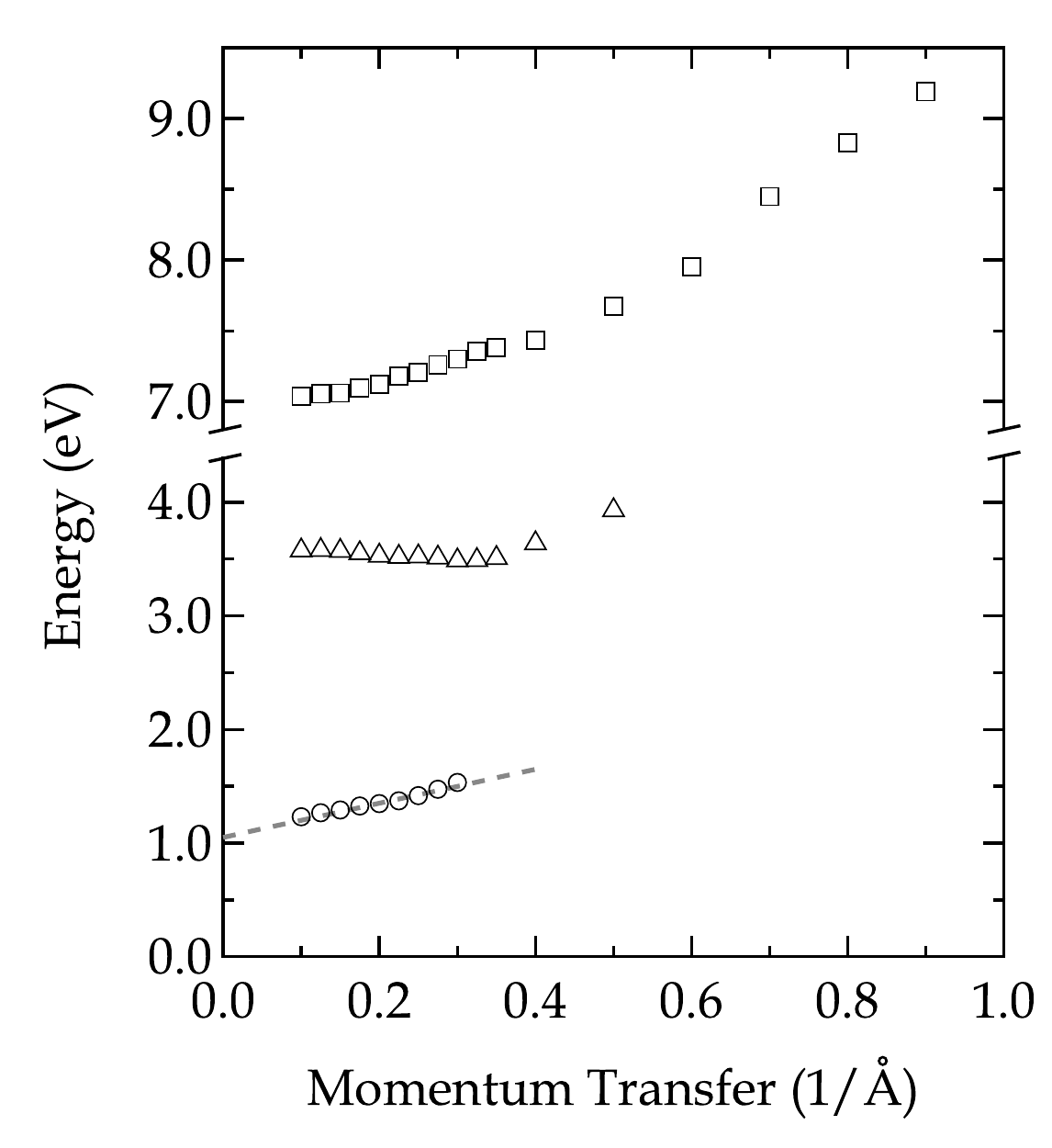}
\caption{Left panel: Momentum dependency of calcium intercalated graphite in a range between 0.1 - 0.35\,\AA$^{-1}$ measured at 20\,K. Right panel: Peak position for the first three maxima as a function of momentum transfer. (The grey dashed lines in both spectra can be seen as a guide for the eye.)}
\label{f5}
\end{figure*}

For CaC$_6$ the modifications of the spectra are more complex. A strong Ca\,3$p$ core excitation feature can be observed at higher energies ($\approx$ 36\,eV). Furthermore, as shown in Fig.\,\ref{f3}, the former $\pi$ plasmon is suppressed and broadened and two new features at about 1.2\,eV (very weak, for further discussion see below and Fig.\,\ref{f5}) and 3.8\,eV occur upon calcium intercalation. Recently, time-dependent density functional based calculations of the loss function of CaC$_6$ have been published \cite{Echeverry2012}, and in Fig.\,\ref{f4} the calculated loss function for small momentum transfers in the $a,b$ crystallographic plane is compared to our experimental result. The agreement is remarkably good, in particular in view of the relatively large energy window as displayed in Fig.\,\ref{f4}. We take this agreement as further evidence for the successful preparation of CaC$_6$.

\par

In Ref.\,\citenum{Echeverry2012} the feature at about 6 - 7\,eV has been assigned to the $\pi$ plasmon of CaC$_6$, which is shifted to lower energies as compared to pristine graphite. This downshift as a function of intercalation parallels the situation in potassium intercalated KC$_8$. The lower energy shoulder and broadening, which is clearly observed for CaC$_6$ signals that the $\pi$ bands are more strongly altered by the addition of Ca as compared to K. This conclusion is in agreement with the formation of hybrid bands in CaC$_6$ and the variation of the C\,1$s$ core excitations as shown above.

\par

Furthermore, the excitation feature at about 3.5\,eV has been termed intra-band plasmon, i.\,e., it is ascribed to the plasmon of the conduction electrons in CaC$_6$. Finally, we observe a low energy maximum at about 1.2\,eV, whereas the calculations predicted a small excitation feature in the loss function at 0.8\,eV. Interestingly, this low energy feature has been ascribed to an acoustic plasmon, a plasmon that is characterized by zero energy at vanishing momentum. The presence of acoustic plasmons in more complex electron systems with, e.\,g., significantly different effective masses of charge carriers has been discussed in the literature previously. The existence of acoustic modes was recently predicted for Pd and MgB$_2$ \cite{Silkin2009,Silkin2009_2}. A very interesting aspect is that such acoustic plasmons in principle are also able to mediate the pairing of charge carriers to Cooper pairs, i.\,e., to a superconducting ground state, and thus can take over the role of phonons in conventional superconductors. For CaC$_6$  however, it has been also argued that acoustic plasmons cannot exist since the different bands present at the Fermi level are neither different enough in terms of their effective masses nor satisfy the condition of being two-dimensional. We will address the character of the lowest energy excitation in CaC$_6$ in the following paragraphs, where we will discuss the momentum dependence (dispersion) of the observed features.

\par

The $\pi$ plasmon is a generic feature of all extended $sp^2$ hybridized carbon system, such as graphite or carbon nanotubes, and it reflects the collective excitation of the $\pi$ system as a whole. The dispersion of the $\pi$ plasmon signals the delocalization of the $\pi$ states in such materials. This remains unchanged for CaC$_6$. This observation is in very good agreement with the \emph{ab initio} time-dependent density functional theory calculations of the dispersion of CaC$_6$ published recently \cite{Echeverry2012}. The calculations also indicate (within the $a$,$b$-plane) a significant dispersion of the $\pi$ plasmon, which is in quantitative agreement to our data (cf.~right panel of Fig.\,\ref{f5}).

\par

Moreover, the appearance of a new, strong feature in the range between 3 and 4\,eV is also a results of the theory, and was interpreted as an intraband plasmon (IP). This IP plasmon first shows a slight negative dispersion before it loses spectral weight due to Landau damping, the decay of the plasmon into individual electron-hole excitations at higher momentum transfers. The initial negative dispersion is in contrast to what would be expected for a simple (Drude-like) charge carrier plasmon, and it has been argued that band structure effects are responsible for this observed behavior. The impact of the particular band structure on the dispersion of charge carrier plasmons, in particular resulting in a negative plasmon dispersion, has also been discussed for other materials \cite{Faraggi2012,Cudazzo2012}. For CaC$_6$, there is even quantitative agreement between the experimental data and the predictions taking into account these band structure effects as regards the sign and size of the dispersion.

Additionally, as mentioned above, a further weak feature centered at 1.2\,eV for $q$\,=\,0.1\,\AA$^{-1}$ can be detected in the electronic
excitation spectrum of CaC$_6$. As shown in Fig.\,\ref{f5}, this feature shifts to higher energies with increasing momentum transfer and the analysis of the peak maximum indicates a linear dispersion. A similar low energy excitation was the result of the DFT calculations, whereas the energy positions are somewhat lower (at $q$\,=\,0.181\,\AA$^{-1}$ the peak was predicted at about 0.8\,eV) \cite{Echeverry2012}. Interestingly, also the theory predicts a quasi-linear dispersion of this excitation.

\par

However, there is an important difference between the theory results and our measurements. An extrapolation of the experimentally obtained
dispersion to $q$\,=\,0\,\AA$^{-1}$ results in an energy value which is different from zero ($\sim$~1\,eV) and thus contradicts the conclusion that this
excitation represents an acoustic plasmon. Furthermore, also a comparison of the slope of the dispersion differs significantly by a factor of $\sim$~3 between experiment and theory. While the overall agreement of the calculated and measured loss function is excellent, the contrasting slopes and optical limits for the lowest energy plasmon do not corroborate the identification of the predicted acoustic plasmon in CaC$_6$. Especially, the non vanishing energy in the optical limit rules out the observed plasmon as potential mediator for the superconducting coupling in CaC$_6$. Our findings are in favor of phonon mediated superconducting coupling. Further work is necessary in order to settle the character of the low energy excitation around 1.2\,eV.


\section{Summary}
We have demonstrated that Ca can be intercalated into graphite (HOPG) under ultra-high vacuum conditions, in order to produce CaC$_6$ thin
films. The electronic excitation spectrum as measured using electron energy-loss spectroscopy changes significantly upon this Ca intercalation. The $\pi$ plasmon shows a downshift in energy, while two new collective excitations appear at 3.5\,eV and 1.2\,eV, respectively. All these features show a dispersion upon increasing momentum transfer. Following recent calculations based on density functional theory, we assign the new feature at 3.5\,eV to an intrabend plasmon, i.\,e., a collective excitation of charge carriers in CaC$_6$. The observed initial negative dispersion is in very good, quantitative agreement to the theoretical predictions and has been assigned to the impact of band structure effects. Concerning the lowest excitation feature at about 1.2\,eV, the experimental data and the calculations however do not agree. Our data do not support the existence of an acoustic plasmon in this energy range.

\begin{acknowledgments}
We thank M. Naumann, R. H\"ubel and S. Leger for technical assistance. This work has been supported by the Deutsche Forschungsgemeinschaft
(Grant No. KN393/13).
\end{acknowledgments}

\end{document}